\documentclass[reprint,amsmath,amssymb,aps]{revtex4-2}

\usepackage{graphicx}
\usepackage{dcolumn}
\usepackage{bm}
\usepackage{hyperref}
\usepackage{lipsum}
\usepackage{subcaption}  
\usepackage{caption}         
\usepackage{float}
\usepackage{tikz}
\usepackage{ragged2e}
\usetikzlibrary{positioning, arrows.meta}  
\usepackage{relsize}  
\usepackage{overpic}  
\usepackage{placeins}
\usepackage{amssymb}
\usepackage{amsmath}
\usepackage[english]{babel}
\usepackage{dblfloatfix}
\usepackage{makecell}
\usepackage{bm}
\usepackage{placeins}
\usepackage[export]{adjustbox}

\begin{document}

\preprint{APS/123-QED}

\title{Spin Freezing in Oscillator Ising Machines:\\ When Second Harmonic Injection Impedes Computation}


\author{Malihe Farasat}
\affiliation{%
University of Virginia, Charlottesville, VA, USA
}%

\author{E.M.H.E.B. Ekanayake}
\affiliation{%
University of Virginia, Charlottesville, VA, USA
}%

\author{Nikhil Shukla}
\affiliation{%
University of Virginia, Charlottesville, VA, USA
}%

\begin{abstract}
Second harmonic injection (SHI) has emerged as a critical mechanism in enabling networks of coupled oscillators to function as Oscillator Ising Machines (OIMs), capable of minimizing the Ising Hamiltonian. While SHI facilitates phase binarization essential for mapping oscillator phases to spin states, we demonstrate that it can also induce a previously unreported phenomenon---spin freezing---where oscillator spins are unable to transition between spin states, even when such a transition can reduce the Ising energy. This freezing effect can impair the analog dynamics of the OIM, preventing it from reaching lower-energy spin configurations. Through theoretical analysis and numerical simulations, we show that the onset of spin freezing is highly sensitive to the initial phase configuration of the oscillators. Contrary to conventional practice, which favors random initialization, we find that initializing all oscillators at specific phase values ($\phi = \pi$ or $\phi = \frac{\pi}{2}$) delays the onset of spin freezing and consistently yields higher-quality solutions. These findings point to the need to carefully engineer the SHI for optimal performance.

\end{abstract}

\maketitle

\section{\label{sec:level1} Introduction}
Combinatorial optimization, as a problem class, remains a formidable challenge for traditional digital computing architectures, providing impetus for the exploration of alternate computational paradigms. One promising direction involves involves engineering physical systems whose intrinsic dynamics naturally minimize the objective functions associated with combinatorial optimization problems (COPs). This paradigm has garnered significant attention, with proposed hardware implementations spanning quantum (e.g., quantum annealers), optical (e.g., coherent Ising machines), spin (e.g., spinwave), and electronic (oscillator Ising machines) domains \cite{mohseni2022ising,zhang2024review,todri2024computing}.

Among these, the Oscillator Ising Machine (OIM) represents a compelling microelectronics platform that is compatible with CMOS-process technology \cite{moy20221,graber2024integrated,mallick2021overcoming,lo2023ising,maher2024cmos}. OIMs leverage the natural dynamics of coupled oscillators to minimize the Ising Hamiltonian:

\[
H = -\sum_{i,j} J_{ij} s_i s_j
\]

\noindent where \(s_i \in \{+1, -1\}\) denotes the spin state and \(J_{ij}\) encodes the interaction between spins \(i\) and \(j\). In typical implementations, OIMs consist of networks of coupled oscillators operating under second harmonic injection (SHI) \cite{Wang2021}. While a detailed quantitative analysis of SHI is presented in the following sections, qualitatively, its role can be understood as enforcing phase binarization i.e., driving the oscillator phases ($\phi$) toward discrete values in the set $\{0, \pi\}$, which can then be mapped to spin states $s\in\{+1, -1\}$, respectively.

Previous studies have primarily focused on the influence of SHI on the local stability of Ising solutions~\cite{bashar2020experimental,cheng2024control}, as well as empirical studies on its potential use as an annealing mechanism~\cite{bashar2021experimental}. However, its impact on the \textit{temporal dynamics of computation} remains underexplored. 

In investigating this dimension, the present work uniquely identifies the phenomenon of \textit{spin freezing}, wherein oscillator spins become unable to transition between spin states—even when such transitions can lower the Ising energy —due to the influence of SHI. We show that spin freezing strongly impacts the system dynamics and the solution quality. Moreover, the emergence of spin freezing in the OIM dynamics is highly sensitive to the choice of the initial conditions. Contrary to conventional expectations—and despite the widespread use of random phase initialization—we find that initializing the oscillators at specific phase values, namely $\phi = \pi$ (or 0), or $\phi=\frac{\pi}{2}$ in the presence of noise, delays the onset of spin freezing and leads to improved solution quality.

\begin{figure}[htbp]
    \centering
    \includegraphics[width=1\linewidth]{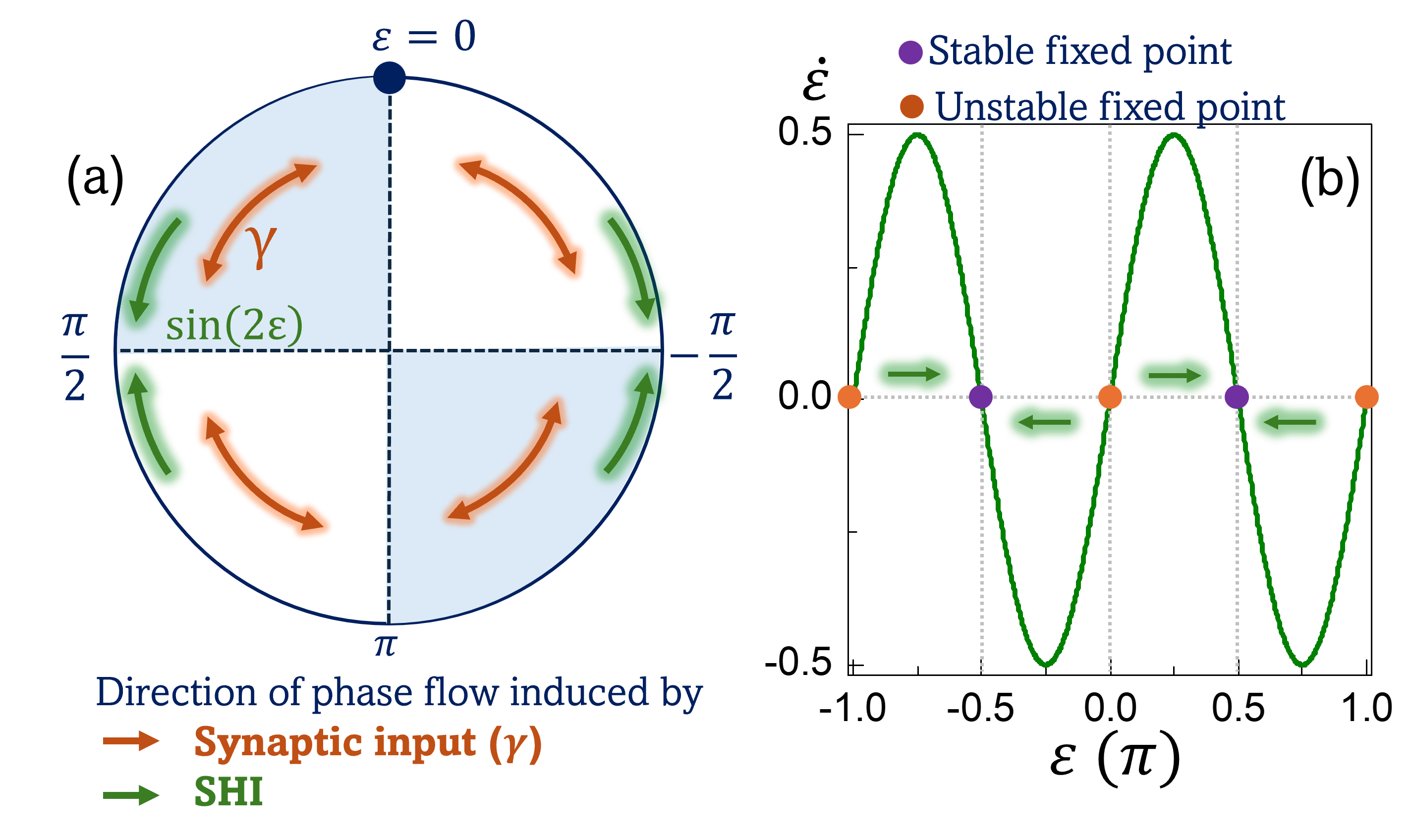}
    \caption{ \textbf{Competing dynamics in OIMs}. (a) Direction of the phase flows generated by the synaptic input ($\gamma$) and the second harmonic injection (SHI) illustrated on a phase plot. (b) Simulated example of the phase flow ($\dot\epsilon$) induced by SHI (only) for $K_s = 0.5$. Here, $\gamma$ is assumed to be zero. The SHI term acts as a restoring force that drives each oscillator phase toward the nearest stable fixed point which represents its current spin state. This restoring behavior can induce spin freezing, where oscillators are unable to flip their spin despite network feedback.}
    \label{fig:fig1}
\end{figure}

\section{oscillator spin freezing induced by SHI}
To illustrate how the SHI influences the OIM dynamics, we begin by examining the behavior of coupled Kuramoto oscillators subjected to SHI as described by,

\begin{equation}
\begin{split}
\frac{d\phi_i}{dt} &= -K \sum_{\substack{j=1 \\ j \ne i}}^N J_{ij} \sin(\phi_i - \phi_j) - K_s \sin(2\phi_i) \label{eq:OIM_dynamics}\\\\
&\equiv \gamma_i- K_s \sin(2\phi_i) \\
\end{split}
\end{equation}

\noindent Here, $\phi_i$ refers to the phase of oscillator $i$. $K$ and $K_s$ represent the coupling strength among the oscillators and the strength of the SHI, respectively. \(\gamma_i=-K \sum_{\substack{j=1 \\ j \ne i}}^N J_{ij} \sin(\phi_i - \phi_j)\) signifies the feedback from the other oscillators which we refer to as  the synaptic input (to oscillator $i$). The $K_s\sin(2\phi_i)$ term on the RHS of Eq.~\eqref{eq:OIM_dynamics} represents the dynamics induced by SHI. The corresponding energy function, minimized by above dynamics \(\frac{dE}{dt} \leq 0\), can be expressed as,
\begin{align}
E({\phi}) &= -K\sum_{\substack{i,j=1 , j \neq i}}^N J_{ij} \cos\big(\phi_i  - \phi_j\big)  \notag \\
&- K_s \sum_{i=1}^N \cos\big(2\phi_i\big) \label{eq:OIM_energy} \end{align}

As elegantly demonstrated by Wang \textit{et al}.~\cite{Wang2021}, under suitable SHI strength, the energy function of the OIM attains a local minimum at \(\phi \in \{0, \pi\}^N\). Moreover, at these specific phase points (only), the OIM energy function corresponds to the energy of discrete spin configurations i.e., the Ising energy. The dynamics in Eq.~\ref{eq:OIM_dynamics} suggest that, upon initialization, the oscillator phases relax toward \(\phi = 0\) or \(\phi = \pi\). Here, we note that $\phi$ is expressed in its wrapped form such that $\phi \in [0,2\pi)$. 

We now turn our attention to analyzing the conditions under which the oscillator phases can traverse the intermediate points, $\phi = \frac{\pi}{2}$, and $\phi = \frac{3\pi}{2}$. Specifically, we examine when a phase $\phi < \frac{\pi}{2}$ can evolve to $\phi > \frac{\pi}{2}$, and vice versa, with analogous considerations for $\phi = \frac{3\pi}{2}$. These intermediate values, referred to here as \emph{neutral points}, are of particular significance for the following reasons:

\begin{itemize}
    \item They represent phase points at which the direction of phase flow induced by SHI reverses, as elaborated in subsequent sections.
    \item For an oscillator spin to flip its state—i.e., transition from $\phi = 0$ ($\pi$) to $\phi = \pi$ ($0$)—it must cross at least one of these neutral points. In many practical implementations, these points serve as thresholds for defining spin states, typically expressed as $s = \text{sgn}\left(\cos(\phi)\right)$.
\end{itemize}

To facilitate this analysis, we rotate the frame of reference by $0.5\pi$, such that $\phi = \frac{\pi}{2} + \epsilon$. This transformation aids in conceptualizing the spin-freezing phenomenon. In the rotated frame, the phase dynamics can be expressed as,

\begin{equation}
\begin{split}
\frac{d\epsilon_i}{dt} &= -K \sum_{\substack{j=1 \\ j \ne i}}^N J_{ij} \sin\left(\frac{\pi}{2} + \epsilon_i - \frac{\pi}{2} - \epsilon_j\right) \\
&\quad - K_s \sin\left(2\left(\frac{\pi}{2} + \epsilon_i\right)\right) \\\\
&= -K \sum_{\substack{j=1 \\ j \ne i}}^N J_{ij} \sin(\epsilon_i - \epsilon_j) + K_s \sin(2\epsilon_i)
\label{eq:rotated_OIM}
\end{split}
\end{equation}

In this transformed coordinate system, the fixed points corresponding to the valid Ising configurations are located at $\epsilon^* \in \left\{-\frac{\pi}{2}, +\frac{\pi}{2}\right\}$. Notably, the SHI term, $K_s\sin(2\epsilon_i)$ vanishes at the neutral points $\epsilon = 0$ (corresponding to $\phi = \frac{\pi}{2}$) and at $\epsilon = \pi$ (corresponding to $\phi = \frac{3\pi}{2}$).

We decompose the dynamics of each oscillator phase $\epsilon_i$ in Eq.~\eqref{eq:rotated_OIM} into two distinct components:

\[
\frac{d\epsilon_i}{dt} = \frac{d\epsilon_i^{(1)}}{dt} + \frac{d\epsilon_i^{(2)}}{dt}
\]

\noindent where,

\[
\frac{d\epsilon_i^{(1)}}{dt} = -K \sum_{\substack{j=1 \\ j \ne i}}^N J_{ij} \sin(\epsilon_i - \epsilon_j), \quad
\frac{d\epsilon_i^{(2)}}{dt} = K_s \sin(2\epsilon_i)
\]

\noindent represent the contributions from the synaptic input and the SHI term, respectively.

Analyzing the SHI component, $K_s \sin(2\epsilon_i)$, we observe the following behavior:

\[
\frac{d\epsilon_i^{(2)}}{dt} =
\begin{cases}
> 0, & \text{if } 0 < \epsilon_i < \frac{\pi}{2} \text{ or } \pi < \epsilon_i < \frac{3\pi}{2} \ (\equiv -\frac{\pi}{2}) \\
< 0, & \text{if } \frac{\pi}{2} < \epsilon_i < \pi \text{ or } -\frac{\pi}{2} < \epsilon_i < 0
\end{cases}
\]

This behavior, illustrated in Fig.~\ref{fig:fig1}(a)(b), indicates that the SHI term acts as a \emph{restoring force}, driving the oscillator phase toward the nearest fixed point. This fixed point corresponds to the oscillator's current spin state. Consequently, the SHI term suppresses transitions between spin states.

\begin{figure*}
    \centering
    \includegraphics[width=0.9\linewidth]{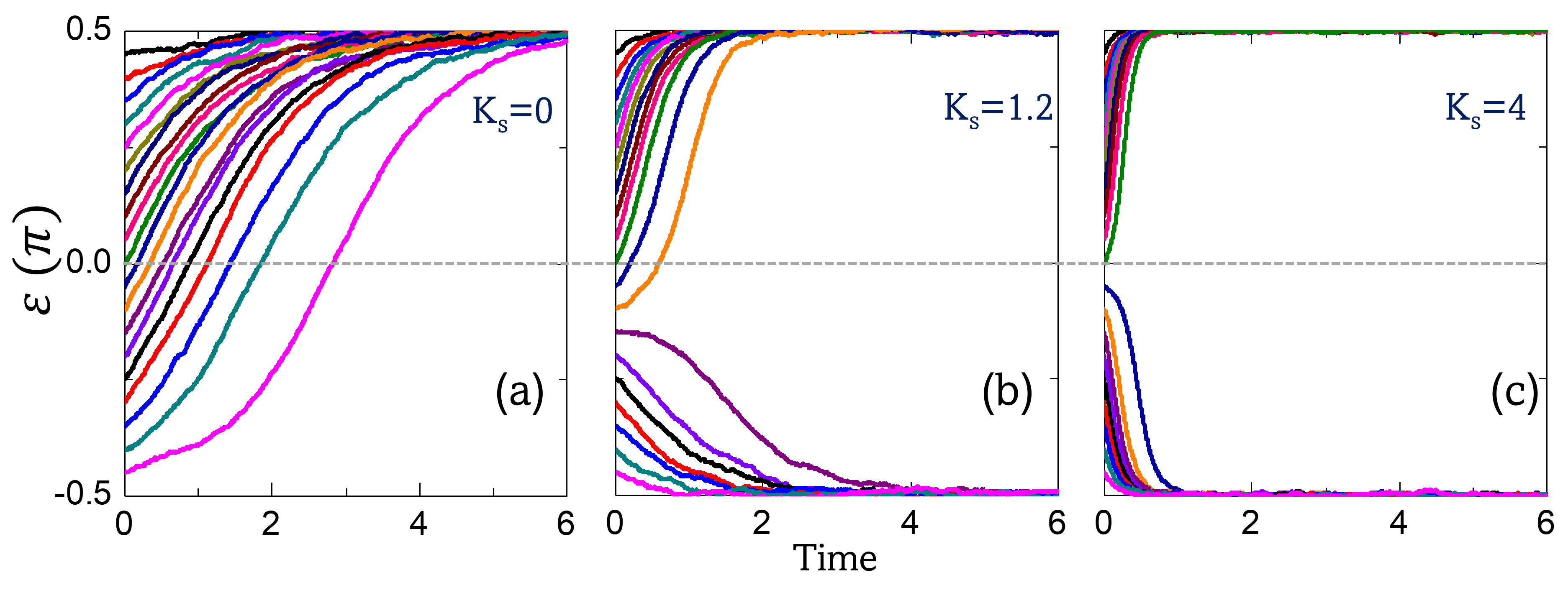}
    \caption{\textbf{Spin freezing in OIMs.} Evolution of the phase of oscillator 1 ($\epsilon_1$) in a negatively coupled two-oscillator network under varying levels of SHI: (a) $K_s = 0$; (b) $K_s = 1.2$; (c) $K_s = 2$. The phase of oscillator 2 is fixed at $\epsilon_2 = -\frac{\pi}{2}$. As $K_s$ increases, $\epsilon_1$ may become frozen and converge to $\epsilon_1 = -\frac{\pi}{2}$, preventing a spin flip even though $\epsilon_1 = -\frac{\pi}{2}$ is not an energetically favorable configuration. This illustrates how SHI can inhibit spin transitions and potentially degrade solution quality.}
    \label{fig:fig.2}
\end{figure*}

In contrast, the term \( \frac{d\epsilon_i^{(1)}}{dt} \), which represents the synaptic feedback from the other oscillators in the network, depends on both the current phase \( \epsilon_i \) and the phases of its connected neighbors, represented by $\epsilon_j$. This component can drive the system either toward or away from its nearest fixed point, and thus, allows the system to change its spin state and reduce the Ising energy. This component is depicted using bi-directional arrows in Fig.~\ref{fig:fig1}(a). We emphasize that only the synaptic input can help change the oscillator spin's state.

Moreover, this decomposition allows us to formulate a condition under which an oscillator spin $i$ becomes effectively \emph{frozen}. Specifically, if

\[
\left|\frac{d\epsilon_i^{(1)}}{dt}\right| < \left|\frac{d\epsilon_i^{(2)}}{dt}\right|,
\]

\noindent then the synaptic input is unable to overcome the restoring influence of the SHI term. Consequently, the oscillator cannot transition to a different spin state effectively inhibiting the minimization of the \emph{Ising energy}. We refer to such a spin as a frozen spin. As noted earlier, the Ising energy refers to the energy associated with the discrete spin states. It is important to note that while a frozen oscillator spin cannot minimize the \emph{Ising energy}, it may still follow local gradient descent and minimize the \emph{energy function of the OIM} (Eq.~\eqref{eq:OIM_energy}), which satisfies the condition \(\frac{dE}{dt}\leq0\).

As discussed further below, spin freezing is most consequential to the computational functionality when \( \frac{d\epsilon_i^{(1)}}{dt} \) and \( \frac{d\epsilon_i^{(2)}}{dt} \) have opposite signs, thereby pushing the dynamics in opposing directions. To ensure that a spin remains \emph{unfrozen} and capable of transitioning state, the SHI strength, $K_s$, must satisfy the following inequality:

\begin{equation}
\begin{split}
    &\quad \quad \left|K \sum_{\substack{j=1 \\ j \ne i}}^N J_{ij} \sin(\epsilon_i - \epsilon_j)\right| > \left|K_s \sin(2\epsilon_i)\right| \\\\
    &\Rightarrow K_s < \left|\frac{K \sum_{\substack{j=1 \\ j \ne i}}^N J_{ij} \sin(\epsilon_i - \epsilon_j)}{\sin(2\epsilon_i)}\right|.
    \label{eq:spin_flip}
\end{split}
\end{equation}
We note that in the presence of noise, the constraint on $K_s$ formulated above becomes probabilistic rather than deterministic, leading to a softened or diffuse boundary for spin freezing.

The freezing of oscillator spins can be conceptualized as being analogous to a reduction in the system’s degrees of freedom, which in turn can degrade its ability to minimize the Ising Hamiltonian effectively. This phenomenon is illustrated with an example discussed below.  

We now illustrate spin freezing with the simple example of a two-oscillator system with negative coupling under the influence of SHI. For simplicity, we fix the phase of oscillator 2 at $\epsilon_2 = -\frac{\pi}{2}$, while initializing the phase of oscillator 1, $\epsilon_1$, from various starting points. We then track the evolution of $\epsilon_1$ across different values of SHI strength, as illustrated in Fig.~\ref{fig:fig.2}. Given the negative coupling between the oscillators, the energetically favorable configuration corresponds to $\epsilon_1 = \frac{\pi}{2}$.

From the observed dynamics, we note that when $K_s = 0 \Rightarrow  \frac{d\epsilon_1^{(2)}}{dt}=0$, the phase of oscillator 1 converges to $\epsilon_1 = \frac{\pi}{2}$ regardless of its initial value. This behavior is expected, as the condition for spin freezing is never satisfied in the absence of SHI. In contrast, when $K_s > 0$, the two components—$\frac{d\epsilon_1^{(1)}}{dt}$ and $\frac{d\epsilon_1^{(2)}}{dt}$—begin to counteract each other when oscillator 1 is initialized at $\epsilon < 0$. Depending on the initial phase and the magnitude of $K_s$, oscillator 1 may become \emph{frozen}, evolving toward $\epsilon_1 = -\frac{\pi}{2}$ even though $\epsilon_1 = \frac{\pi}{2}$ represents a spin configuration with lower Ising energy. As \(K_s\) increases, the range of  phase values that lead to spin freezing also expands. For instance, for \(K_s = 1.2\), oscillator 1, when initialized at \(\epsilon_1 = \{-0.05\pi, -0.1\pi\}\), can still flip its spin state. However, no such transition is observed at the same phase initialization for \(K_s = 4\). In this example, we also observe that when \(\epsilon > 0\), the synaptic input and SHI reinforce each other. Thus, even if spin freezing occurs, it does not adversely affect the computational outcome, since both the components drive oscillator 1 toward \(\epsilon_1 = \frac{\pi}{2}\), the lowest-energy spin configuration.

\begin{figure*}[htbp]
    \centering
    \includegraphics[width=0.8\linewidth]{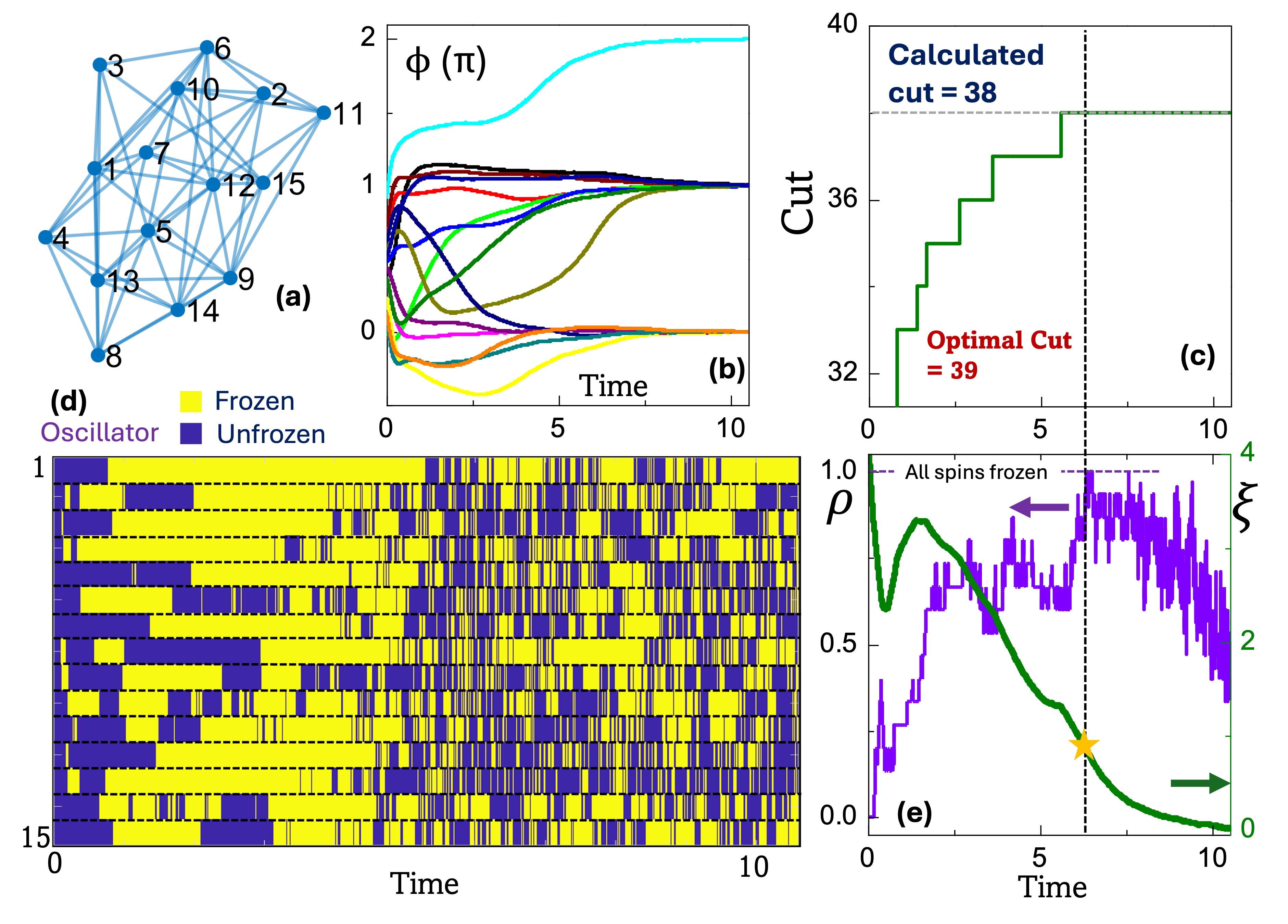}
    \caption{\textbf{Impact of spin freezing on temporal dynamics of OIMs.} (a) Illustrative 15-node graph used in the simulation. Temporal evolution of: (b) oscillator phases ($\phi$);  (c) computed graph cut; (d) spin state of each oscillator (yellow: frozen, blue: unfrozen); (e) fraction of frozen spins ($\rho$) and phase deviation parameter ($\xi$). The results show that the emergence and progressive increase of spin freezing can effectively impede computation even before the system has achieved steady state. $K=1;\:K_s=1 \text{ and } K_n=0.005$ are used in the simulation.}
    \label{fig:fig.3}
\end{figure*}

\begin{figure*}[htbp]
    \centering
    \includegraphics[width=1\linewidth]{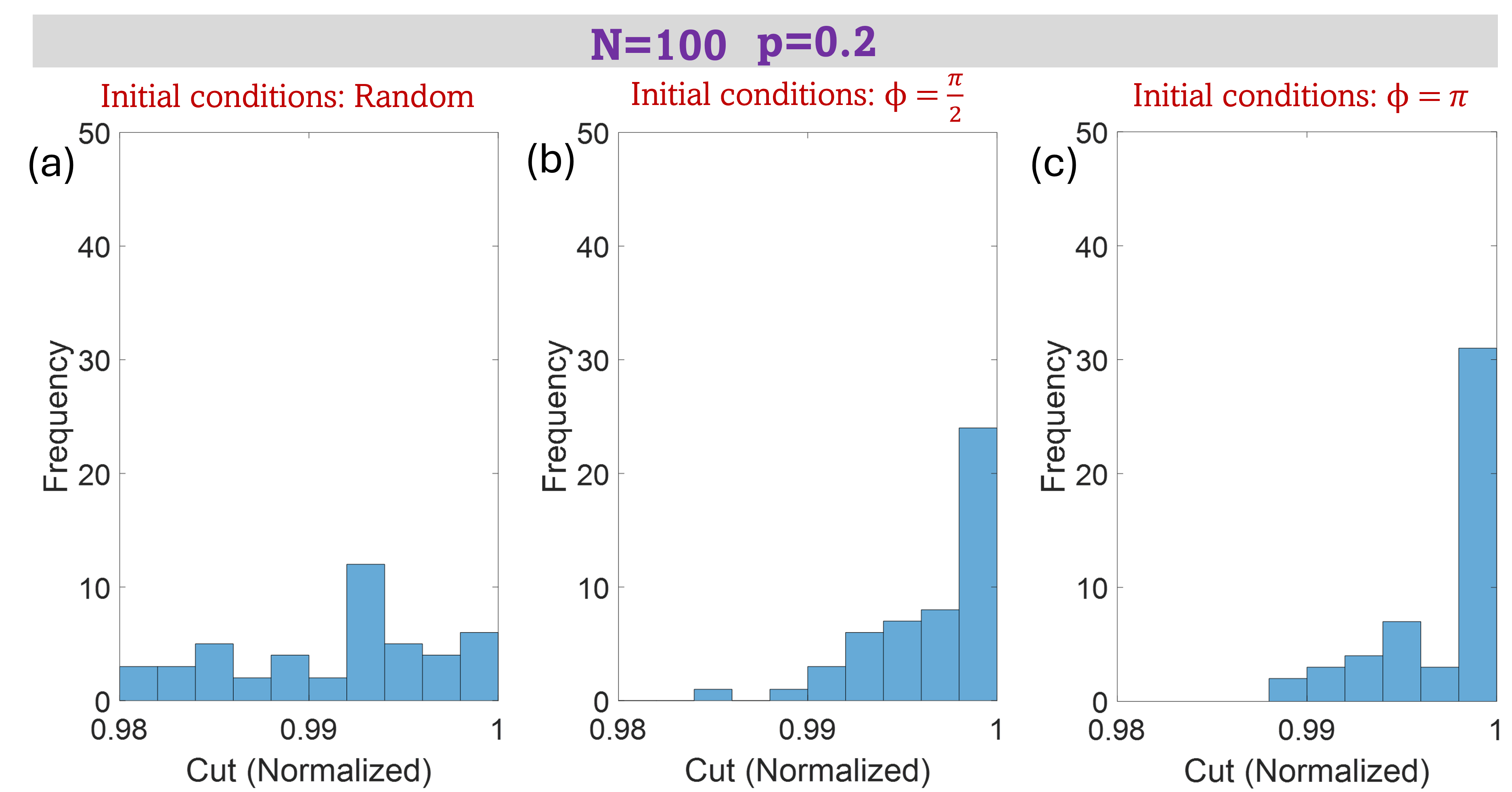}
    \caption{\textbf{Impact of initial conditions on solution quality.} Histograms of the normalized graph cut obtained under different initialization strategies for 100-node Erd\H{o}s--R\'enyi graphs with edge probability $p = 0.2$: (a) oscillator phases initialized randomly; (b) all oscillators initialized to $\phi = \frac{\pi}{2}$; (c) all oscillators initialized to $\phi = \pi$. Initializing the oscillator phases to $\phi = \frac{\pi}{2}$ or $\phi = \pi$ consistently yields higher-quality solutions. Similar trends are observed across graphs of varying sizes, $V = \{100, 150, 200\}$ nodes and edge probability, $p=\{0.2, 0.4, 0.5, 0.6, 0.8\}$, as shown in Appendix~\ref{appendix2}.}
    \label{fig:fig.4}
\end{figure*}

\section{Impact of spin freezing on temporal dynamics of OIM}
We now investigate using simulations, the impact of spin freezing in OIM networks. We note that although a rotated frame of reference was employed earlier to elucidate the phenomenon of spin freezing, all simulation results presented henceforth are reported in the original phase frame ($\phi$) to maintain consistency with the conventions commonly adopted in literature.

Our analysis focuses on the MaxCut problem, which involves partitioning the nodes of a graph into two disjoint sets such that the total weight of the edges crossing the partition is maximized. This problem maps directly to the anti-ferromagnetic Ising model, where \( J_{ij} = -W_{ij} \). Here, \( W_{ij} \) and \( J_{ij} \) represent the weights of the edges in the original graph and the corresponding spin interaction graph (solved by the OIM), respectively. The Ising Hamiltonian \( H \) is related to the MaxCut value as, \(H = \sum_{i,j}W_{ij}-2 \cdot \text{MaxCut}\). Thus, computing the MaxCut is equivalent to finding the ground state of the Ising Hamiltonian.

\begin{figure*}[htbp]
    \centering
    \includegraphics[width=1\linewidth]{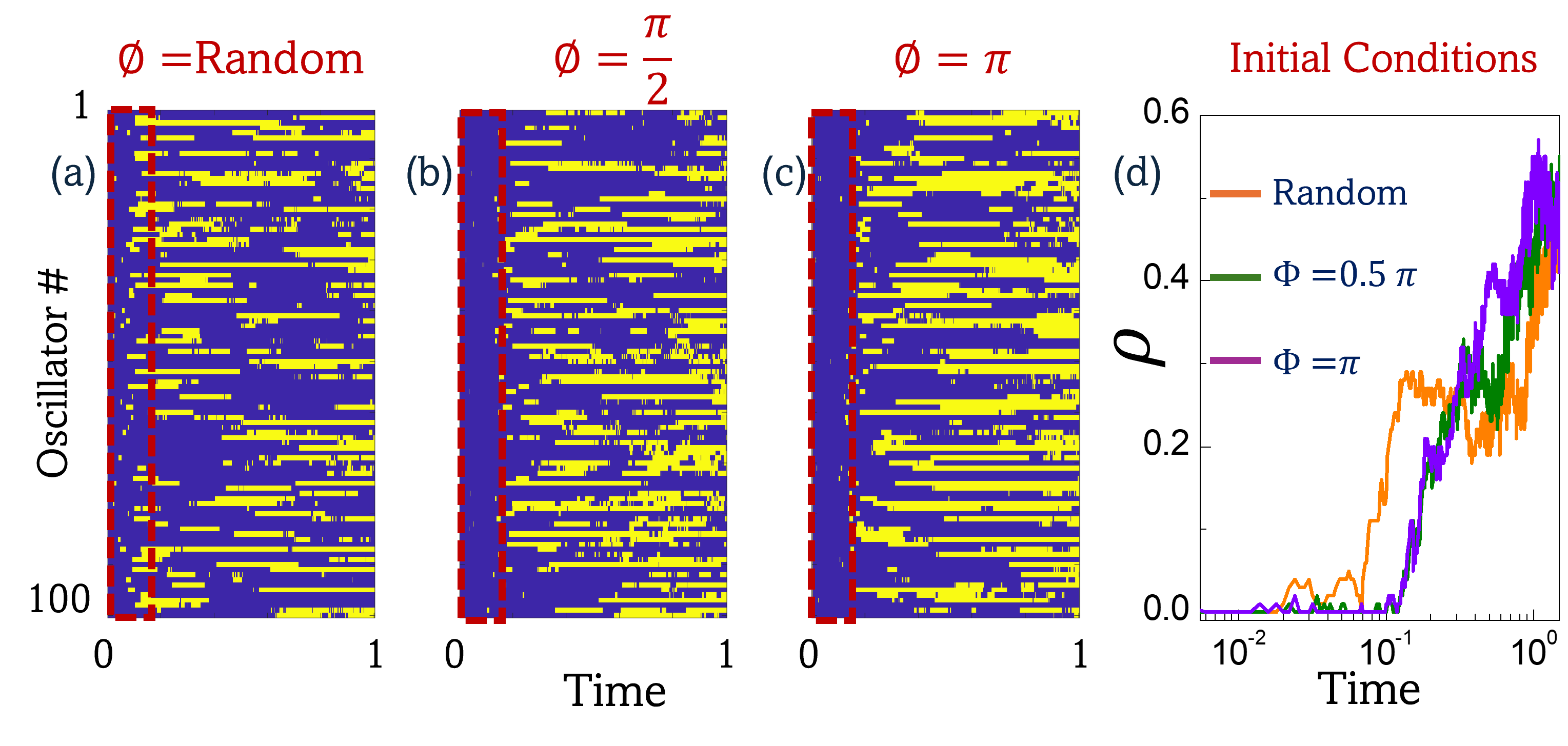}
    \caption{\textbf{Impact of initialization strategy on the onset of spin freezing.} Heatmaps showing the spin state of oscillators under different initialization strategies: (a) random phase initialization; (b) all oscillators initialized to $\phi = \frac{\pi}{2}$; (c) all oscillators initialized to $\phi = \pi$. (d) Temporal evolution of the fraction of frozen spins for each initialization condition. Initializing the oscillator phases to $\phi = \frac{\pi}{2}$ or $\phi = \pi$ delays the onset of spin freezing allowing the system more time to switch spin configurations.}
    \label{fig:fig.5}
\end{figure*}

We begin with an illustrative example of a small, randomly generated graph consisting of 15 nodes and 58 edges, shown in Fig.~\ref{fig:fig.3}(a). We simulate the oscillator dynamics and monitor, at each time step, both the fraction of oscillators that are \emph{frozen}- i.e., unable to switch spin states- and the corresponding graph cut. The graph cut is calculated by the thresholding the phase, $ s = \text{sgn} \left( \cos\left(\phi\right)\right)$. The simulations are performed using a stochastic differential equation solver implemented in \textsc{MATLAB}\textsuperscript{\textregistered}. For this example, the simulation is performed by initializing the oscillator phases to random values; however, the impact of different initialization strategies will be explored in the following section.

Figures~\ref{fig:fig.3}(b) and (c) illustrate the evolution of oscillator phases and the temporal progression of the graph cut, respectively. Figure~\ref{fig:fig.3}(d) presents a heatmap depicting the instantaneous state of each oscillator spin—\textit{yellow} indicates frozen spins, while \textit{blue} denotes unfrozen ones. It is important to note that, due to the dynamic nature of the system, a spin classified as frozen at a given time instant may subsequently unfreeze. However, as illustrated in the example considered earlier, spin freezing--- even when intermittent--- can adversely affect the system's ability to descend towards lower energy spin configurations.

Figure~\ref{fig:fig.3}(e) presents the temporal evolution of two key metrics: the fraction of frozen spins, \( \rho \), and the phase deviation from the nearest fixed point, \( \xi \), given by:

\[
\xi  = \pi \sum_{i=1}^N \left| \frac{\vartheta_i}{\pi}  -  \frac {\left( 1-\text{sgn}\left(\cos\left(\vartheta_i\right)\right)\right)}{2} \right| 
\]

\noindent where,

\[
\vartheta_i = \min\big(\phi_i \bmod 2\pi,\: 2\pi - \left(\phi_i \bmod 2\pi\right)\big)
\]

\noindent is defined such that it accounts for the periodic nature of the oscillator phase. This formulation ensures that the deviation is always measured with respect to the nearest fixed point in the phase space.

The evolution of \( \rho \) in Fig.~\ref{fig:fig.3}(e) shows that the system initially exhibits a relatively small (but not 0) proportion of frozen spins, also evident in the heatmap in Fig.~\ref{fig:fig.3}(d). However, $\rho$ rapidly increases to over 70\% by $T=3$. Furthermore, at \( T \sim 6 \), the system state is characterized by all the spins being frozen. When all the spins have frozen, the OIM effectively ceases computation for that duration. Notably, the system has not yet reached steady-state since \( \xi \approx 1\). No further improvement in the graph cut is observed even though the cut is sub-optimal in nature. The system computes a graph cut of 38 while the optimal MaxCut is 39.

We note that in the presence of noise, the system may not converge precisely to the $\phi \in \{0,\pi\}$ fixed points. Therefore, we define the 'steady state' as the condition where oscillator phases approach either \( \phi = 0 \) or \( \phi = \pi \), as seen in the phase trajectories (Fig.~\ref{fig:fig.3}(b)), and the deviation parameter \( \xi \) plateaus to a small value (Fig.~\ref{fig:fig.3}(e)). In the present example, the system reaches steady state at approximately \(T \sim 10\). This result demonstrates that spin freezing, induced by SHI, can cause the system to cease computation—or enter periods of inactivity (in terms of performing computation)—even before reaching steady state. Here, we emphasize that computation refers to the transition of oscillator spin states aimed at minimizing the Ising energy.

\section{Influence of Initial Conditions on SHI-Driven OIM}
Next, we investigate how  initialization conditions impact the spin freezing behavior in OIMs and their resulting computational properties. To evaluate this, we first generate  Erdos-Renyi graphs with different sizes, $V=\{100, 150, 200\}$ nodes and different edge probability, $p=\{0.2, 0.4, 0.5, 0.6, 0.8\}$. We consider 50 graphs for every combination of $V$ and $p$, and compute their MaxCut using the OIM dynamics. Each graph is simulated 10 times and the best solution from the 10 runs is recorded as the computed MaxCut result for that graph. This process is performed for three sets of initialization conditions:

\begin{itemize}
    \item Oscillators initialized to random phase values  \(\phi=[0,\pi]\).
    \item All oscillators initialized to \(\phi=\frac{\pi}{2}\).
    \item All oscillators initialized to \(\phi=\pi\). Considering the inherent symmetry of the system, this is equivalent to initializing all the oscillators to \(\phi=0\). 
\end{itemize}

\noindent The simulation parameters have been detailed in Appendix \ref{appendix1}. Figure~\ref{fig:fig.4} presents histograms of the normalized graph cut for the case \( V = 100 \), \( p = 0.2 \), under three different initialization conditions. Results for all other cases are provided in  Appendix~\ref{appendix2} and exhibit similar trends in terms of the  as the impact of the initial conditions on the solution quality. Each MaxCut solution is normalized to $\theta=\max(P,Q)$, where $P:$ the best cut obtained across all runs and initialization conditions for the corresponding graph, and $Q$: cut obtained from Gibbs sampling after $10^6$ sweeps. For 150 and 200 node graphs shown in Appendix~\ref{appendix2}, the number of sweeps is increased to $3\times 10^6$ and $5\times10^6$, respectively.

The results presented in Fig. \ref{fig:fig.4} as well as those shown in the appendix show that initializing the oscillators in the OIM at \( \phi = \frac{\pi}{2} \) or \( \phi = \pi \) more consistently yields larger cut values, or, in other words, lower Ising energies. No specific trend is observed between these two initialization strategies. In contrast, random initialization results in the lowest probability of achieving such cuts. This finding is particularly counterintuitive, given the prevalence of random initialization in conventional implementations of these systems.

To explore the underlying cause behind the strong influence of initial conditions, we analyze the fraction of frozen spins and its temporal evolution across the three initialization scenarios. A representative example for a 100-node graph is shown in Figs.~\ref{fig:fig.5}(a–c), which display heatmaps of the instantaneous spin states of the oscillators; additional examples have been presented in Appendix~\ref{appendix3}. Figure~\ref{fig:fig.5}(d) shows the evolution of the fraction of frozen spins over time for the three initialization scenarios. Interestingly, for initializations at \(\phi = \frac{\pi}{2}\) and \(\phi = \pi\), spin freezing begins noticeably later compared to the random initialization case, where some spins freeze early in the system's dynamical evolution. This early onset of frozen spins suggests that the gradient descent dynamics are more strongly influenced by SHI, which inhibits spin transitions and impairs the system’s ability to effectively minimize the Ising energy. As a result, the \(\phi = \frac{\pi}{2}\) and \(\phi = \pi\) initialization strategies yield better statistical performance compared to random initialization.

\section{Conclusion}
The results presented in this work reveal that the SHI strength, $K_s$, plays a dual role in OIMs: while it is essential for enforcing phase binarization and stabilizing spin states, it can also inhibit spin state transitions and induce spin freezing. Notably, $K_s$ (relative to $K$) can be viewed as being analogous to the inverse temperature in Boltzmann statistics, where higher values suppress configurational changes and reduce the system's ability to explore the solution space.

Spin freezing has significant implications for both the performance and operational behavior of OIMs. Our results demonstrate that the onset of spin freezing is strongly influenced by the initial phase configuration of the oscillators. Consequently, careful engineering of SHI parameters is critical for optimizing OIM performance. While this study focused on constant $K_s$ and $K$, incorporating temporal variations—such as annealing schedules —may provide a pathway to mitigate some of the adverse effects of spin freezing and improve solution quality \cite{albertsson2023highly,Wang2021,bashar2021experimental}. Additionally, recent work on heterogeneous SHI injection \cite{allibhoy2025global}, where each oscillator is assigned a different $K_s$, can also be a potentially promising direction.

Finally, our findings motivate further exploration of spin freezing in emerging OIM variants, including those with higher-order interactions \cite{bashar2023designing,bybee2023efficient,bashar2023oscillator}, dynamical Ising machines \cite{cdc9-y234}, K-state oscillator Potts machines \cite{mallick2022computational}, and Lagrange-based oscillator networks \cite{delacour2025lagrange}. Understanding and managing spin freezing in these systems may play an important role in optimizing their computational behavior and overall performance.

\section*{Acknowledgments}
This material was based upon work supported by the National Science Foundation (NSF) under Grant No. 2328961 and was supported in part by funds from federal agency and industry partners as specified in the Future of Semiconductors (FuSe) program.\\

\textbf{Author Contributions:} \textbf {Malihe Farasat:} Conceptualization (equal);  Writing – review \& editing (equal). {E.M.H.E.B Ekanayake:} Conceptualization (equal); Writing – original draft (equal); Writing – review \& editing (equal). \textbf {Nikhil Shukla:} Conceptualization (lead); Funding acquisition (lead); Supervision (lead); Validation (equal); Writing – original draft (lead); Writing – review \& editing (lead). \\

\textbf{DATA AVAILABILITY:} The data that support the findings of this study are available from the corresponding author upon reasonable request.

\appendix
\section{\MakeUppercase{Simulation Parameters}}
\label{appendix1}

To simulate the OIM dynamics in the presence of noise, we express Eq. \eqref{eq:OIM_dynamics}  as,

\begin{align}
{d}\phi_i
&= \Bigl[
   -K \sum_{\substack{j=1,j\neq i}}^{N} J_{ij}\,\sin\bigl(\phi_i-\phi_j\bigr) \nonumber \\
&\quad 
   -K_s\,\sin\bigl(2\phi_i\bigr)
  \Bigr]\,{d}t +\,K_n\,{d}W_t\,, \label{appendix:SDE2}
\end{align}

where is $K_n$ represents the noise power. Equation \eqref{appendix:SDE2} is subsequently simulated using a standard SDE solver implemented in \textsc{MATLAB}\textsuperscript{\textregistered}. The parameters used in the simulation for computing the the graph cuts on the Erd\H{o}s--R\'enyi graphs are shown in \ref{tab:table1}.

\renewcommand{\arraystretch}{1.2}
\begin{table}[htbp]
\centering
\scriptsize 
\resizebox{0.8\columnwidth}{!}{%
\begin{tabular}{|c|c|c|c|c|}
\hline
\textbf{$N$} & \textbf{$K$} & \makecell{\bm{$K_s$}} & \makecell{\bm{$K_n$}} & \makecell{\textbf{MCMC} \\ \textbf{sweeps}} \\
\hline
100 & 1 & 1 & 0.05 & $1\times 10^6$ \\
\hline
150 & 1 & 1.5 & 0.06 & $3\times 10^6$ \\
\hline
200 & 2 & 3 & 0.08 & $5\times 10^6$ \\
\hline
\end{tabular}
}
\caption{Parameters used in the simulation of the graph cuts on the Erd\H{o}s--R\'enyi graphs.}
\label{tab:table1}
\end{table}

\section{\MakeUppercase{Impact of Initial Conditions on Solution Quality}}
\label{appendix2}

Histograms of the normalized graph cut values computed for Erdős–Rényi random graphs of varying sizes, $V = \{100,\: 150,\: 200\}$, and edge probabilities, $p = \{0.2, 0.4, 0.5, 0.6, 0.8\}$, initialized from different starting conditions, are presented in Fig.~\ref{fig:fig.6c}(a--c).

\begin{figure*}
    \centering
    \includegraphics[width=1\linewidth]{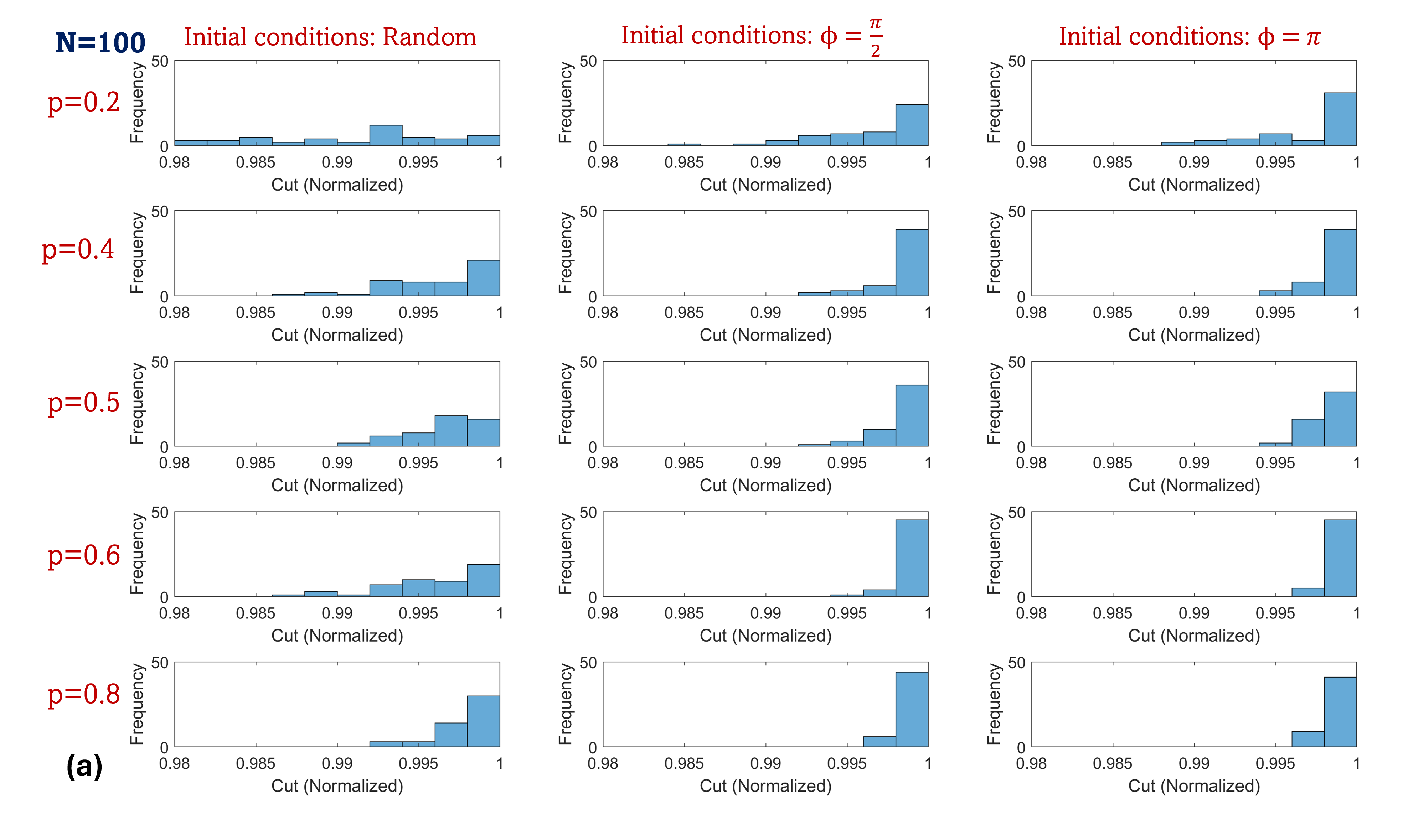}
\end{figure*}

\begin{figure*}
    \centering
    \includegraphics[width=1\linewidth]{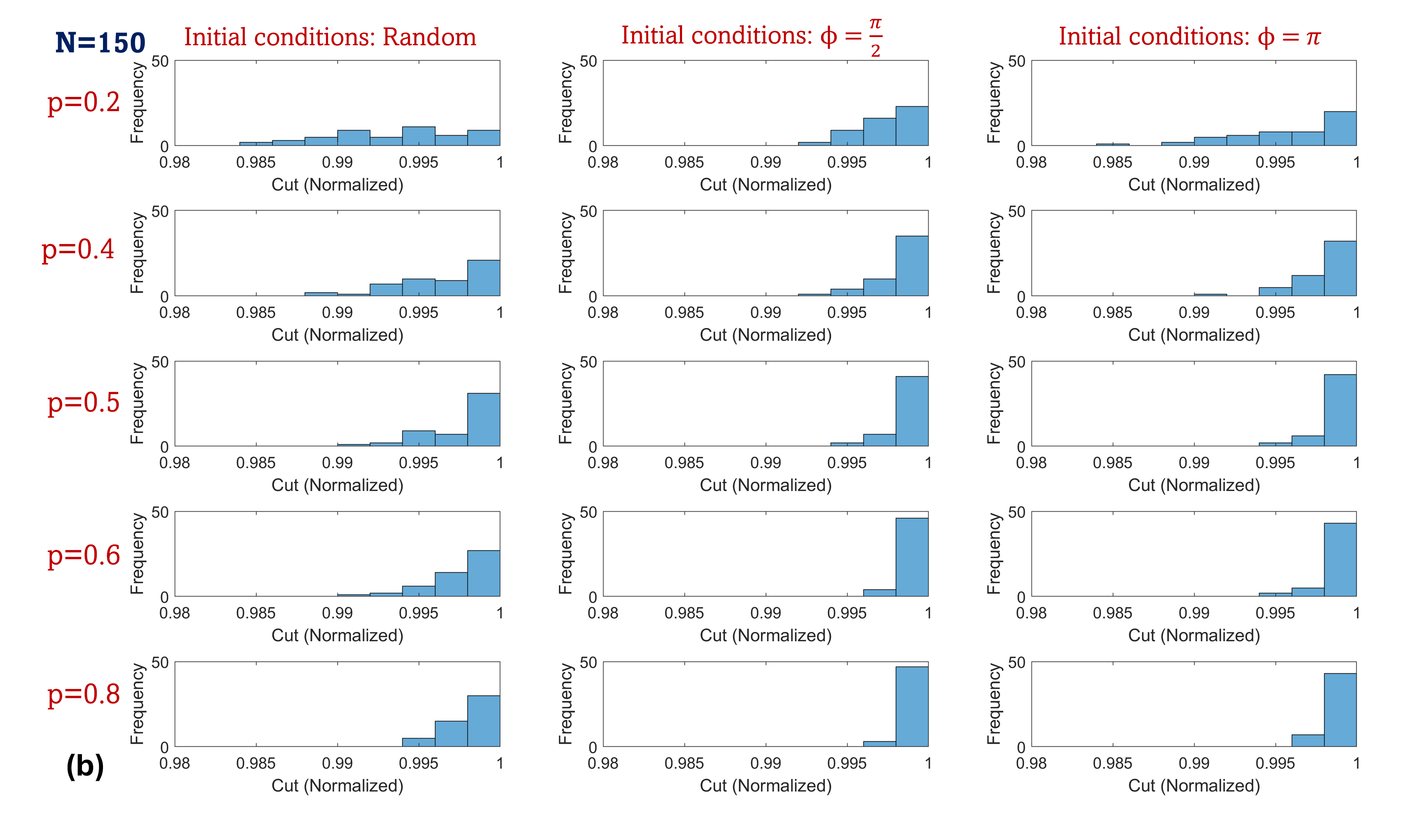}
\end{figure*}

\begin{figure*}
    \centering
    \includegraphics[width=1\linewidth]{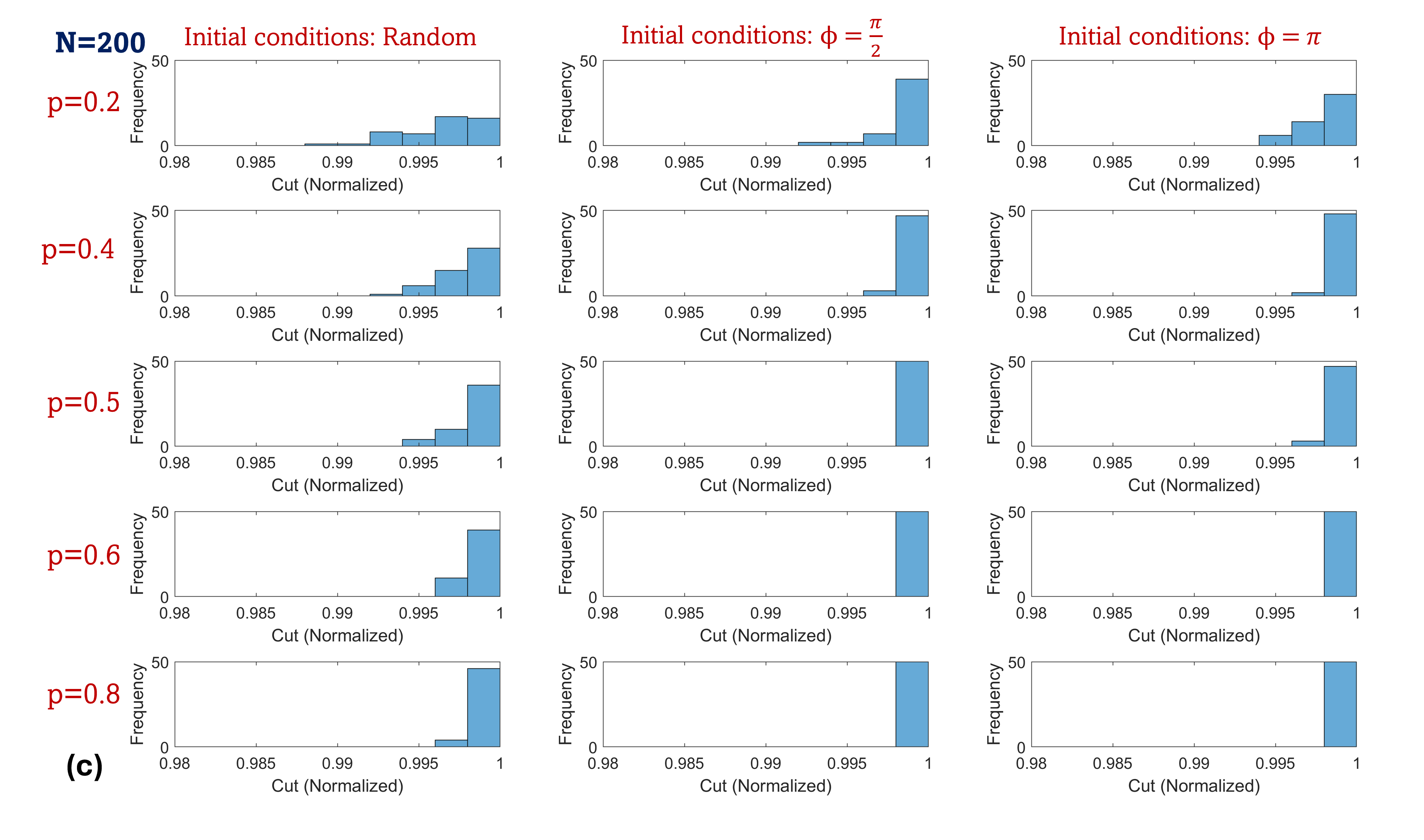}
    \caption{\textbf{Impact of initialization conditions on spin freezing.} Histograms of the normalized graph cut obtained under three different initialization strategies across randomly generated Erdős–Rényi graphs of varying sizes ($N$) and edge probabilities, $p = \{0.2, 0.4, 0.5, 0.6, 0.8\}$: (a) $N = 100$; (b) $N = 150$; (c) $N = 200$. The results show that initializing oscillator phases to $\phi = \frac{\pi}{2}$ or $\phi = \pi$ consistently yields better solution quality by delaying the onset of spin freezing.}
    \label{fig:fig.6c}
\end{figure*}

\clearpage
\section{\MakeUppercase{Onset of Spin freezing}}
\label{appendix3}

\begin{figure}[!h]
    \centering
    \includegraphics[width=1\linewidth]{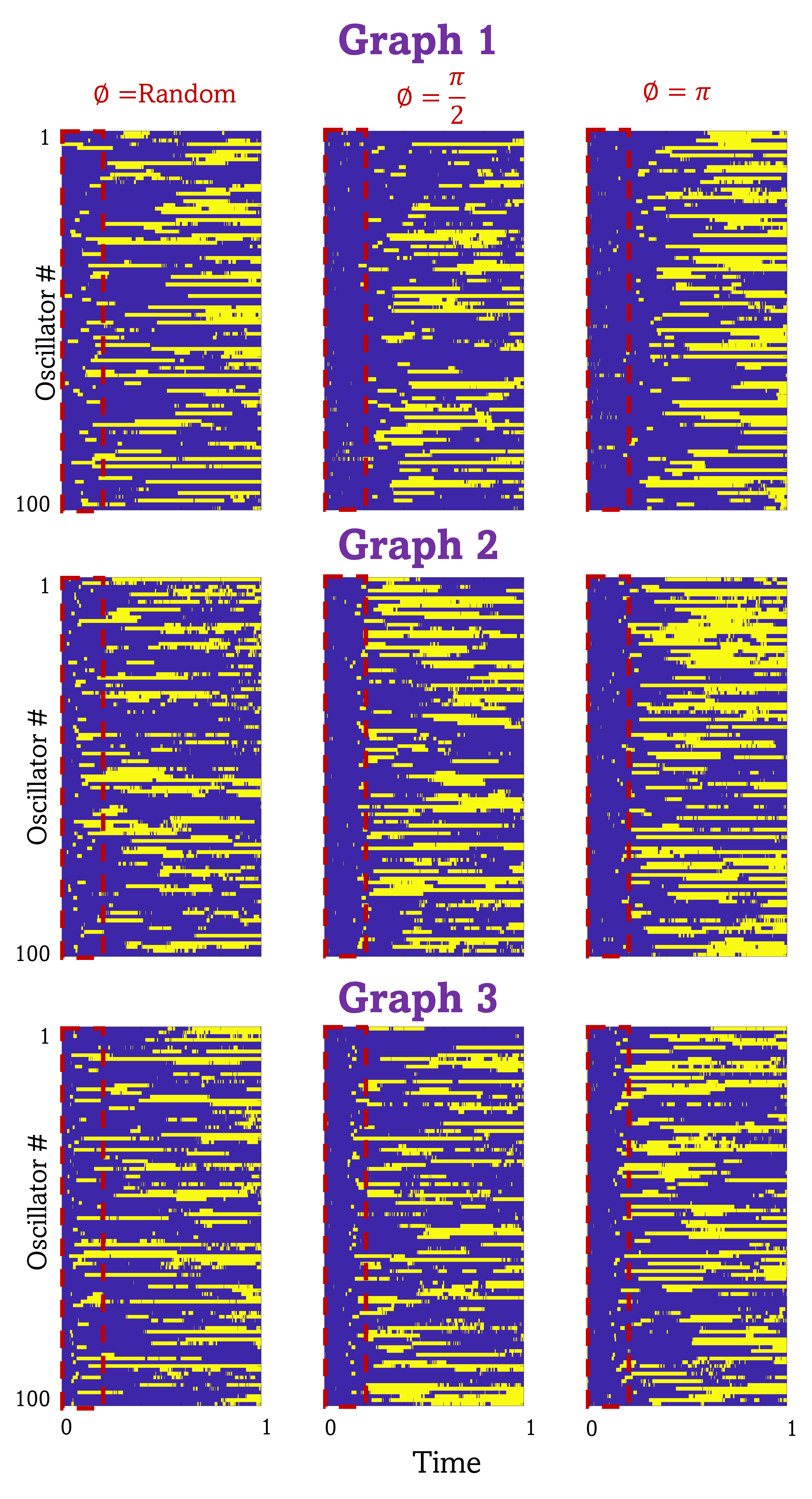}
     \caption{\textbf{Impact of initialization strategy on the onset of spin freezing.} Heatmaps for 3 randomly generated 100-node graphs showing the spin state (yellow: frozen; blue: unfrozen) of oscillators under different initialization strategies: (a) random phase initialization; (b) all oscillators initialized to $\phi = \frac{\pi}{2}$; (c) all oscillators initialized to $\phi = \pi$.}
    \label{fig:fig.7}
\end{figure}

We consider additional graphs to show how the onset of spin freezing is influenced by different initialization strategies. Figure~\ref{fig:fig.7} shows heatmaps for 3 randomly generated graphs showing the spin state (yellow: frozen; blue: unfrozen) of oscillators under different initialization strategies.

\def\bibsection{\section*{References}}  
\bibliography{main.bbl}  
\end{document}